\documentclass[aps,prl,preprint,amssymb,amsmath]{revtex4-1}

\usepackage{graphicx}%
\usepackage{dcolumn}%
\usepackage[colorlinks=true,linkcolor=blue]{hyperref}
\usepackage[dvipsnames]{xcolor}

\begin{document}

\title{Band Structure for Thermal Conduction in Multilayered Systems. Thermal Crystals.}

\author{A. Camacho de la Rosa, D. Becerril$^{\dagger}$, G. G\'omez-Farf\'an, R. Esquivel-Sirvent}
\email{raul@fisica.unam.mx}
\affiliation{Instituto de F\'{i}sica, Universidad Nacional Aut\'onoma de M\'exico, Apartado Postal 20-364, Ciudad de M\'{e}xico, 01000, Mexico.\\
$^{\dagger}$ Istituto di Struttura della Materia, CNR, Via Fosso del Cavaliere 100, 00133 Rome, Italy. }

\date{\today}

\begin{abstract}
In this paper we solve the Cattaneo-Vernotte Equation for a periodic heterostructure made of alternate layers of different materials. The solutions describe thermal waves traveling in a periodic system, and it allows us to introduce the concept of thermal crystals. We show that the dispersion relation shows the characteristics of a band-structure, however the corresponding Bloch wave vector is always complex corresponding to pseudo-bands, unlike what happens in photonic or acoustic crystals. In this context, we also discuss the use of the Floquet-Bloch theorem for thermal waves.  The case of finite layered structures is also analyzed showing the possibility of changing the temperature and heat flux by introducing defects opening the possibility of thermal management through the pseudo-band structure. 

\end{abstract}

\maketitle

\section{Introduction}
Band-gap engineering has become an essential method for tuning or changing transport properties in materials. Stacking of layers of different dielectric functions leads to photonic crystals, and the transmission and reflection properties are controlled by a judicious choice of the materials and the thickness of the individual layers \cite{Yariv,John,YUE201545}.
The transmission properties of a photonic crystal can be changed externally, for example, applying mechanical stress to change the spacing of the layers in what is called tunable mirrors \cite{Porta,JENA2019113627}, or using Mott transitions \cite{Rashidi}. The transmission bands change by choosing a combination of metamaterials with a negative index of refraction \cite{Shokri,Sy} or intercalating graphene layers to construct omnidirectional photonic crystals \cite{ALSHEQEFI2015127}. Alternate layers of metals and dielectrics permit the transmission of light over a tunable range of frequencies, such as the ultraviolet, the visible, or the infrared wavelength range \cite{Scalora}. Band-structure engineering is also applied to induce topological transitions \cite{Rudner}. Similar results are obtained for acoustic crystals where the acoustic impedance replaces the dielectric function. The periodicity of the structure gives a band-structure that for acoustic transmission \cite{Manvir,Cocol,SiC,Lai}. The existence of photonic and acoustic band structures is based on Bloch-Floquet theorem \cite{Harrison,Cottey} used for periodic structures. 

Heat conduction, as described by Fourier's law, can also be changed by layered media. Depending on whether the heat flow is parallel or perpendicular to the layers, system's thermal conductivity is modeled by an equivalent series or parallel resistance array \cite{Libro}. If the layers have an anisotropic thermal conductivity, the heat flow can be redirected by changing the thermal anisotropy of successive layers \cite{Vemuri,Vemuri2}. In superlattices of Si/Ge the thermal transport can also be changed by a mechanical strain \cite{BORCATASCIUC2000199}. Minimizing the thermal conductivity in periodic metal lattices is now possible \cite{Chen}. 

Unlike the quantum, photonic and acoustic cases which are use a wave equation (hyperbolic equation), the Fourier law follows a parabolic partial differential equation and, as a diffusive system, does not exhibit a band type structure in the heat flux.  Furthermore, Fourier's heat law implies an infinite heat propagation velocity.  To account for a finite velocity of heat propagation, a correction introduced by Cattaneo extended Fourier theory to a hyperbolic equation \cite{hyperbolic}. The wave like behavior derived from Cattaneo's equation opens new possibilities for thermal management. Its wave behavior has been discussed by several authors \cite{Salazar_2006, OSTOJASTARZEWSKI2009807,KUNDU20123030}, and possible applications include the design of thermal metamaterials where it is possible to have thermal cloaking \cite{Alu}. Shendeleva \cite{Shendeleva} studied thermal wave reflection and refraction at an interface between two media, showing that it is possible to have  an analog to the total reflection angle present in wave theory. The thermal behavior of skin also follows a non-Fourier behavior and this has been experimentally observed, also is employed in the study of bioheat transfer mechanisms  \cite{XU20082237,KOVACS2015613,AHMADIKIA}. Cattaneo equation is not Lorentz invariant, however for most applications this is not an issue. Discussions about its limitations  and certain misconceptions regarding the Cattaneo equations have been discussed in the literature \cite{Spigler,Bright}.
 
In this paper, we solve the Cattaneo equation for a periodic layered medium system (superlattice). This is not the first work to treat this problem. Chen \citep{chen2018heat} calculated the dispersion relation and showed that applying Bloch's theorem a band structure is obtained. However, there is no further discussion regarding the applicability of Bloch's theorem and, as we show, the condition for a band-structure is not satisfied and we can talk of a pseudo-band structures.

 In the first section, we introduce the  Cattaneo equation and the appropriate boundary conditions. In Section II, the matrix formalism for a layered system is derived. In Section IV results will be discussed for some case studies. Finally, conclusions and outlook will be presented.

\section{Model of heat conduction} 

Fourier's Law (FL) of heat conduction is one of the most famous equations in physics, it is a phenomenological description  lacking a fundamental deduction. FL establishes that  heat transfers  from a higher to a lower temperature region, through the linear relation \cite{1993heat,Libro,OSTOJASTARZEWSKI2009807}

\begin{equation}
\vec{q}(\vec{x},t)=-\kappa\nabla T(\vec{x},t),
\label{eq:FL1}
\end{equation}

\noindent
where $\vec{q}$ is the heat flux, $T$ is the temperature and $\kappa$ is the thermal conductivity. In the context of linear response theory, the thermal conductivity can be seen as a proportionality constant between a perturbation of the thermal field described by the temperature gradient, and the system response i.e. the  heat flux $\vec{q}$. However, as can be seen from Eq.~\eqref{eq:FL1} FL assumes an instantaneous  linear response of the system, which would mean that the thermal perturbation has an infinite propagation velocity \cite{1993heat}. 

 Cattaneo and Vernotte independently proposed the insertion of a delay time $\tau$ in the response of a heat flux which is generated by a temperature gradient. This modification to FL  imposes a causal linear response in the heat conduction \citep{sellitto}, which leads to a finite propagation velocity of the thermal perturbation.  Eq.~\eqref{eq:FL1} becomes
 
\begin{equation}
\vec{q}(\vec{x},t+\tau)=-\kappa\nabla T(\vec{x},t).
\label{eq:CV1}
\end{equation}

Assuming a small delay time and carrying out a first order approximation in $\tau$ gives a new heat conduction equation

\begin{equation}
\vec{q}(\vec{x},t)+\tau\dfrac{\vec{q}(\vec{x},t)}{\partial t}=-\kappa \nabla T(\vec{x},t).
\label{eq:CV2}
\end{equation}

The delay time $\tau$, depends on the material and it is in the range $10^{-3}$ s to $10^{3}$ s as was suggested by \citep{luikov}. We transform Eq.~\eqref{eq:CV2} into a Partial Differential Equation (PDE) of the temperature by considering the balance equation of the thermal energy density $u(\vec{x},t)$ given by

\begin{equation}
\dfrac{\partial u(\vec{x},t)}{\partial t}+\nabla\cdot\vec{q}(\vec{x},t)=s(\vec{x},t)
\label{eq:BalanceEnergy}
\end{equation} 

\noindent
with $s$ a source term. When the system does not exchange matter and is held at constant volume, the energy density satisfies 

\begin{equation}
\dfrac{\partial u(\vec{x},t)}{\partial t}=C_{v}\dfrac{\partial T(\vec{x},t)}{\partial t},
\label{eq:ThermalEnergy}
\end{equation}

\noindent
where $C_{v}$ is the heat capacity at constant volume. Applying the divergence to Eq.~\eqref{eq:FL1} and Eq.~\eqref{eq:CV2} in the absence of sources we obtain

\begin{equation}
\nabla^{2}T(\vec{x},t)=\dfrac{1}{\alpha}\dfrac{\partial T(\vec{x},t)}{\partial t}
\label{eq:FL2}
\end{equation}

\noindent
and

\begin{equation}
\nabla^{2}T(\vec{x},t)=\dfrac{1}{\alpha}\dfrac{\partial T(\vec{x},t)}{\partial t}+\dfrac{1}{\upsilon^{2}}\dfrac{\partial^{2} T(\vec{x},t)}{\partial t^{2}},
\label{eq:CV3}
\end{equation}

\noindent
where we have defined the thermal diffusivity $\alpha=\kappa/C_{v}$, and $\upsilon^{2}=\alpha/\tau$. 

Eq.~\eqref{eq:FL2} is the PDE associated to FL, it is a parabolic equation and its solutions describe a diffusion process \citep{tijonov}.  Eq.~\eqref{eq:CV3} is known as the Hyperbolic Heat Conduction Equation or the Cattaneo-Vernotte Equation (CVE). The  parameter $\upsilon$ is identified as the propagation speed of the thermal perturbation \cite{herrera,barletta}. In the limit $\tau\rightarrow 0$,  speed $\upsilon$ goes to infinity, and Eq.~\eqref{eq:CV3} tends to Eq.~\eqref{eq:FL2}. 

The CVE describes the heat conduction as a wave phenomena and its solutions are better known as \textit{thermal waves} \cite{joseph}.  In Appendix A, we present a discussion regarding the nature of thermal waves and the difference with other wave phenomena such as electrodynamics. 

To further explore the wave like-nature of CVE we solve Eq.~\eqref{eq:CV3} in Cartesian coordinates  for one dimension, this is 

\begin{equation}
\dfrac{\partial^{2} T(x,t)}{\partial x^{2}}=\dfrac{1}{\alpha}\dfrac{\partial T(x,t)}{\partial t}+\dfrac{1}{\upsilon^{2}}\dfrac{\partial^{2} T(x,t)}{\partial t^{2}}.
\label{eq:layeri}
\end{equation} 

We assume a solution of the form $T(x,t)=e^{-i \omega t} \hat{T}(x)$, being $\omega$  the angular frequency, leading to 

\begin{equation}
\dfrac{d^{2}\hat{T}}{dx^{2}}= -k^{2}(\omega) \hat{T}(x),
\label{eq:OrdinaryCV}
\end{equation} 

\noindent
where 

\begin{equation}
k^{2}=\dfrac{\omega}{\alpha}\left( i+\dfrac{\omega}{\omega^{*}}\right),
\label{wavevector}
\end{equation}

\noindent
and  $\omega^{*}=\frac{1}{\tau}$. Using the definition of thermal speed propagation $\upsilon$ Eq.~\ref{eq:layeri} , the previous equation can be written as 

\begin{equation}
k^{2}=\dfrac{\omega^2}{\upsilon^2}\left( 1+i\dfrac{\omega^{*}}{\omega}\right),
\label{wavevector2}
\end{equation}

\noindent
that has the usual form of the wave number for a dispersive and dissipative media. 

Thus,  general solutions of Eq.~\eqref{eq:OrdinaryCV}  are

\begin{equation}
\hat{T}(x)=Ae^{ikx}+Be^{-ikx},
\label{eq:solT}
\end{equation}

\noindent
with $A$ and $B$ constants that are determined by the boundary conditions, which are the continuity (i) of the temperature and (ii) of the heat flux that are a consequence of assuming perfect thermal contact. 
 
Equation ~\eqref{wavevector2} is a wavevector, however even for the 1-D solution presented here it is complex for all  frequencies. This is an important difference with  acoustic or electromagnetic wave equations.

Assuming the same frequency dependence in the heat flux $q(x,t)=e^{-i \omega t}\hat{q}(x)$ we have

\begin{equation}
\hat{q}=-s(\omega) \dfrac{d\hat{T}}{dx}=-is k \left(Ae^{ikx}-Be^{-ikx}\right)
\label{eq:solq}
\end{equation}

\noindent
with $s(\omega)=\dfrac{\kappa}{1-i\omega\tau}$. We define $Z(\omega)=s(\omega) k $  as the \textit{thermal impedance}. This quantity describes the opposition of a material to the heat flux when a temperature gradient exists.  Solutions of Eq.~\eqref{eq:layeri} are wave-like, however as pointed out by Salazar \cite{Salazar_2006}
 the time average of the thermal energy is zero $<u(x,t)>=0$, thus the wave carries no energy and  concepts like reflection or transmission must be handled with care. Some previous works directly carry over wave concepts and properties, and calculate quantities such as thermal cloaking \cite{Alu,LIU2014621,XU20082237}. In the Appendix A we show that wave properties are well defined for the Cattaneo-Vernotte equation.

\section{Layered system}
In this section we derive the transfer matrix and the reflectance coefficients for a layered media. The first part is similar to the derivation of Chen \cite{chen2018heat}. 
\begin{figure}[h]
	\centering
	\includegraphics[scale=0.31]{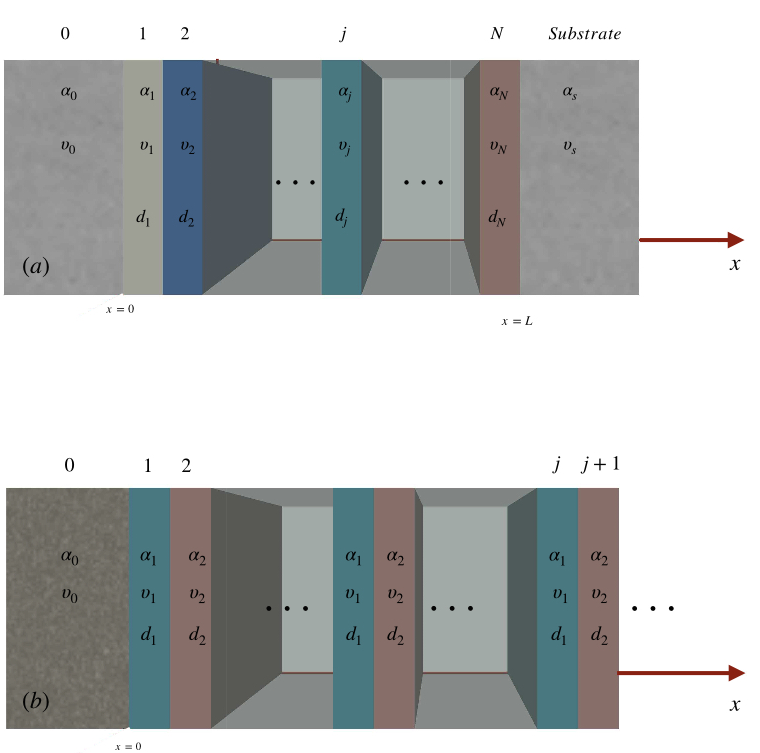}
	\caption{(a) Finite layered system made of consecutive layers with different thermal properties. Each layer is characterized by its thickness $d$, thermal speed $\upsilon$ and the thermal diffusivity $\alpha$. The system is bounded by a substrate and outer medium. (b) Semi-infinite system. The substrate is removed and the layered system extends indefinitely. }
	\label{fig:layeredsystem}
\end{figure}

Consider an array of $N$ layers in a Cartesian coordinates system $(x,y,z)$ where the $x$-axis is perpendicular to the interfaces.  Each layer is of width $d_{j}$ with thermal properties $\alpha_{j},\upsilon_{j}$ as  shown in  Figure \ref{fig:layeredsystem}(a). The first interface is at $x_0=0$ and the last layer interface is at $x_N=d_1+\cdots+d_{N}=L$.  Outside the layered structure, for $x<x_{0}$ there is a medium with thermal properties $\alpha_{0},\upsilon_{0}$; and for $x>x_N$ there is a substrate defined by $\alpha_{s},\upsilon_{s}$.

In a  $j^{th}$ layer,  Equations (\eqref{eq:solT}) and (\eqref{eq:solq}) can be written in a matrix form as

\begin{equation}
\left(\begin{array}{c}
\hat{T}(x)\\
\hat{q}(x)\\
\end{array}\right)=\left(\begin{array}{cc}
1 & 1\\
-iZ_{j} & i Z_{j}\\
\end{array}\right)\left(\begin{array}{cc}
e^{ik_{j} x} & 0\\
0 & e^{-i k_{j} x}\\
\end{array}\right)\left(\begin{array}{c}
A_j\\
B_j\\
\end{array}\right),
\label{eq:matrixproblem}
\end{equation}

\noindent
where we define the non-singular matrix $m_{j}$ as

\begin{equation}
m_{j}(x)=\left(\begin{array}{cc}
1 & 1\\
-iZ_{j} & i Z_{j}\\
\end{array}\right)\left(\begin{array}{cc}
e^{ik_{j} x} & 0\\
0 & e^{-i k_{j} x}\\
\end{array}\right).
\label{eq:minorm}
\end{equation}

The  interfaces of the layer  are at  $x = x_j$ and  $x = x_j + d_j$. The  value of vector  $\vec{T}(x)=\left( \hat{T}(x),  \hat{q}(x)\right)$ evaluated within layer $j$ on the left boundary $x=x_{j}$ is related to the value on the right boundary $x=x_{j}+d_{j}$ by a $2\times 2$ matrix. This is

\begin{equation}
\left(\begin{array}{c}
\hat{T}(x)\\
\hat{q}(x)\\
\end{array}\right)_{x_{j}+d_{j}}=M_{j}\left(\begin{array}{c}
\hat{T}(x)\\
\hat{q}(x)\\
\end{array}\right)_{x_{j}},
\label{eq: matrixjth}
\end{equation}

\noindent
where matrix $M_{j}$ is given by \citep{Cocol}

\begin{equation}
M_{j}=\left(\begin{array}{cc}
\cos (k_{j}d_{j}) & -\frac{1}{Z_{j}}\sin (k_{j}d_{j})\\
Z_{j}\sin (k_{j}d_{j}) &\cos (k_{j}d_{j}) \\
\end{array}\right). 
\end{equation}

Where the applied boundary conditions are the continuity of temperature and heat flux. Thus, at an interface between layers $j$ and $j+1$ we can relate the values of $\vec{T}$ in the different layers as:

\begin{equation}
\left(\begin{array}{c}
\hat{T}(x)\\
\hat{q}(x)\\
\end{array}\right)_{x_{j}+d_{j}+d_{j+1}}=M_{j+1}M_{j}\left(\begin{array}{c}
\hat{T}(x)\\
\hat{q}(x)\\
\end{array}\right)_{x_{j}}.
\end{equation} 

This procedure can be repeated along the whole lattice with the starting point $x_{0}$ and the final point in $x_{N}$. We then obtain

\begin{equation}
\left(\begin{array}{c}
\hat{T}(L)\\
\hat{q}(L)\\
\end{array}\right)=\left( M_{N}M_{N-1}...M_{1}\right)\left(\begin{array}{c}
\hat{T}(0)\\
\hat{q}(0)\\
\end{array}\right).
\label{eq:TransferMatrix}
\end{equation}

The matrix $M=\left( \prod_{j=0}^{N-1}M_{N-j}\right)$ in Eq.~\eqref{eq:TransferMatrix} is known as the Associated Transfer Matrix (ATM), which transfers the fields, in this case temperature and heat flux, through a given domain \citep{perez2004}. 

Using the matrix $m_{j}$ in Eq.~\eqref{eq:minorm} to write $\vec{T}(0)$ and $\vec{T}(L)$ in terms of its coefficients we get

\begin{equation}
m_{N}(L)\left(\begin{array}{c}
A_{N}\\
B_{N}\\
\end{array}\right)=Mm_{1}(0)\left(\begin{array}{c}
A_{1}\\
B_{1}\\
\end{array}\right),
\end{equation}

\noindent
multiplying both sides by $(m_{N}(L))^{-1}$ we obtain

\begin{equation}
\left(\begin{array}{c}
A_{N}\\
B_{N}\\
\end{array}\right)=\left((m_{N}(L))^{-1}Mm_{1}(0)\right)\left(\begin{array}{c}
A_{1}\\
B_{1}\\
\end{array}\right).
\label{eq:ctm}
\end{equation}

The matrix $K=\left((m_{N}(L))^{-1}Mm_{1}(0)\right)$ is known as the Coefficient Transfer Matrix (CTM) \citep{rolando1985}, it transfers the information of the solution from the first to the last layer using the coefficient of the solutions.

We use the CTM to determine the reflectivity $r$ and transmitivity $t$ of the system. We assume a normalization of the temperature multiplying by $A^{-1}_{1}$,  thus the coefficient that describes the original wave incoming into the system is equal to the unity; then $B_{1}/A_{1}\equiv r$ is the fraction of the incoming \textit{thermal wave} that is reflected from the whole structure; likewise $A_{N}/A_{1}\equiv t$  corresponds to the fraction of the \textit{thermal wave} that is transmitted by the whole system; and we assume $B_{N}/A_{1}\equiv 0$, which means that a thermal wave is incident only from one side. Thus, from Eq.~\eqref{eq:ctm}

\begin{equation}
\left(\begin{array}{c}
t\\
0\\
\end{array}\right)=K
\left(\begin{array}{c}
1\\
r\\
\end{array}\right),
\label{eq:ref&trans}
\end{equation}

and we obtain $r$ and $t$ in terms of the components of the matrix $K$ as

\begin{equation}
r=-\dfrac{K_{21}}{K_{22}},
\end{equation}

\noindent
and

\begin{equation}
t=K_{11}-\dfrac{K_{12}K_{21}}{K_{22}}.
\end{equation}

Are obtained the reflectance $R=rr^{*}$, the transmittance $T=tt^{*}$ and the absorbance ${\cal A}$ that satisfy the energy relation 

\begin{equation}
{\cal A}=1-(R+T).
\end{equation}

\subsection{Infinite periodic structure}

Consider an infinite periodic system which can be obtained by translating a unit cell  of  $p$-layers through the spatial period $p=d_{1}+\cdots +d_{p}$ as shown in Fig.~\ref{fig:layeredsystem}(b). In this case we use Floquet's Theorem to study the problem of heat conduction in periodic systems. 

Using the vector  $\vec{T}(x)=(\hat{T},\hat{q})$ and the solutions Eqs.\eqref{eq:solT}-\eqref{eq:solq}, we can write  Eq.~\eqref{eq:OrdinaryCV} as  a first order differential equation  so that for a given layer $j$ we have 

\begin{equation}\label{eq:Floquet1}
\dfrac{d}{dx}\vec{T}(x)=\left( \begin{array}{cc}
0 & -1/s_{j} \\
k_{j}^{2}/s_{j} & 0\\
\end{array}\right) \vec{T}(x)
\end{equation}

\noindent
or  $\frac{d}{dx}\vec{T}(x)=A\vec{T}(x)$. Matrix  $A$ is periodic such that  $A(x+p)=A(x)$ for all $x\geq 0$. The two linearly independent solutions of Eq.~\eqref{eq:Floquet1} define the fundamental matrix $F(x)=(\vec{T}_{1},\vec{T}_{2})$. The Floquet Theorem is used to determine this fundamental matrix, in Appendix B we discuss the differences between Bloch and Floquet as well as the implications of having  a complex wave vector for all frequencies.

Therefore, the propagation of \textit{thermal waves} in a periodic layered system satisfies the expression

\begin{equation}
\left( \begin{array}{c}
\hat{T}(Np)\\
\hat{q}(Np)\\
\end{array}\right)=e^{ iNp {\mathcal Q}}\left( \begin{array}{c}
\hat{T}(0)\\
\hat{q}(0)\\
\end{array}\right).
\label{eq:FloquetCV}
\end{equation}

This result is widely known in solid state physics as Bloch's Theorem \cite{grosso,rolando1985}, which states that the energy eigenstates $\psi(\vec{x})$ for a particle in a crystal characterized by a potential $V(x)=V(x+p)$ of period $p$, are given by 

\begin{equation}
\psi(\vec{x})=e^{i\vec{{\mathcal Q}}\cdot\vec{x}}v(\vec{x})
\end{equation}

\noindent
where $v(\vec{x})\in \mathbb{R}$ is a  function of period $p$, and the wave vector $\vec{{\mathcal Q}}$ is the Bloch vector. We see how Floquet exponents and the Bloch vector play the same role, being the eigenvalues of the translation operator in a periodic system. But the use of Floquet's Theorem allows us to consider \textit{thermal wave vectors} with complex values.

Applying Floquet's theorem to the periodic lattice assuming the total length of the lattice $L$  is  $N$ times the period of each unit cell $p$ we obtain

\begin{equation}
\left( e^{ iNp{\mathcal Q}} \mathbb{I}-M\right)\left( \begin{array}{c}
\hat{T}(0)\\
\hat{q}(0)\\
\end{array}\right)=0,
\label{eq:BlochEquation}
\end{equation}

\noindent
that has a solution only if $\det\left( e^{ iNp{\mathcal Q}} \mathbb{I}-M\right)=0$. We transform this into a  second degree equation

\begin{equation}
e^{2iNp{\mathcal Q}}-e^{iNp{\mathcal Q}}TrM+detM=0,
\label{eq:2nddegree}
\end{equation}

\noindent
where $\det M=1$. We note that \ textit {Floquet's thermal wave vector} is complex, so it is decomposed into its real and imaginary parts, ${\mathcal Q}={\mathcal Q}'+i{\mathcal Q}''$, where both parts must satisfy the following equation

\begin{equation}
\cos(Np({\mathcal Q}'+i{\mathcal Q}''))=\dfrac{1}{2}Tr(M).
\label{eq:bandequation}
\end{equation}

Since we have a wave propagating for $x>0$, the meaningful solutions are those that satisfy $|e^{\mathcal Qd}|<1$. We point out that the values of ${\mathcal Q}$ that satisfy the two resulting implicit equations, define the \textit{thermal dispersion relation} which creates a band structure as would happen in the problem of a particle in a crystal studied with Bloch's Theorem. Care must be taken in the interpretation of the thermal bands. In solid state physics for a  ${\mathcal Q}\in \mathbb{R}$ the energy is real and is associated to a state of the particle within the crystal, but when ${\mathcal Q}$ is purely imaginary, the energy is complex and has no physical meaning so that corresponds to a forbidden band.  The band structure in the thermal case must be interpreted in a different way because ${\mathcal Q}$ is always complex. 

To calculate the reflection of the thermal wave from the semi-infinite structure -Fig.\ref{fig:layeredsystem}(b)- we assume that the interface of the outer medium with impedance $Z_0$ and the layered medium is at $x=0$.  Using Eq.\eqref{eq:BlochEquation} we solve for $r=B_1/A_1|_{x=0}$ to obtain

\begin{equation}
r=\frac{Z_0-Z_{eff}}{Z_{0}+Z_{eff}},
\label{eq:ref-semi}
\end{equation}
where the effective surface impedance of the layered system is defined by 

\begin{equation}
Z_{eff}=\frac{i(e^{i{\mathcal Q}p}-M_{11})}{M_{12}}.
\label{eq:efimped}
\end{equation}

\section{Results and Discussion}

In this section we numerically solve the CVE of a layered system. The parameters used were taken from Chen \citep{chen2018heat} and are in Table \ref{tab:thermalprop}. We consider an infinite periodic system with a unit cell made of two types of materials $1-2$, with period $p=d_{1}+d_{2}$ as shown in Figure \ref{fig:layeredsystem}(b).

\begin{table}[h]
\begin{tabular}{|c|c|c|c|c|}
\hline 
Material & Thermal & Heat & Mass & Response\\ 
           & conductivity   & capacity  & density  & time  \\ 
             & $\kappa$ [$W/mK$]  &$C_{V}$ [$J/K$] & $\rho$ [$1/Kg$] & $\tau$ [$s$] \\ 
\hline 
Material 1 & 0.235 & 3600 & 1500 & 1\\ 
\hline 
Material 2 & 0.445 & 3300 & 1116 & 20\\ 
\hline 
\end{tabular} 
\caption{Thermal properties of the materials used in the numerical example of a layered system taken from \citep{chen2018heat}.}
\label{tab:thermalprop}
\end{table}

\begin{figure}[h]
\centering
\includegraphics[scale=0.1]{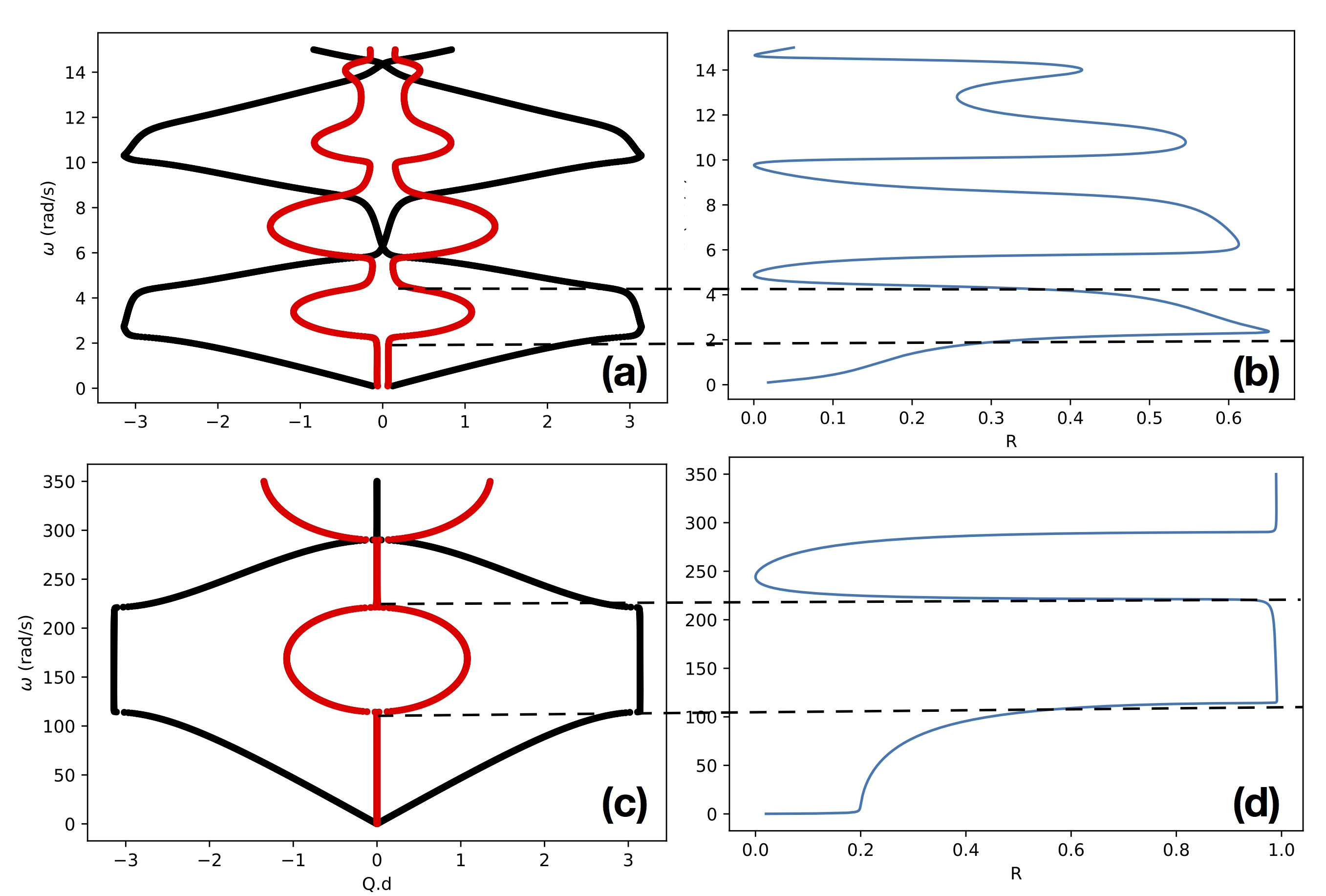}
\caption{In panels (a,c) we plot  the thermal band structures of the semi-infinite periodic layered systems with unit cell $1-2$ and parameters from the Table \ref{tab:thermalprop}. In  (b,d)  the reflectances of the same systems are presented.}
\label{fig:bandstructure}
\end{figure}

We begin by computing the dispersion relation  and reflectance -Eq. \eqref{eq:ref-semi}- for two periodic systems, the first one with layer thickness $d_1=d_2=50$ $\mu m$ -Fig.\ref{fig:bandstructure}(a,b)-, and the second one with layers of thickness $d_1=d_2=1$ $\mu m$ -Fig.\ref{fig:bandstructure}(c,d)-. 
In panels (a) and (c), the imaginary (red line) and real part (black line), $\mathcal{Q}'$ and $\mathcal{Q}''$ are plotted as a function of frequency.  For the structure with a unit cell of period $d=100$ $\mu m$, the dispersion relation (a) resembles a band-structure typical of 1D systems, and the dispersion is bounded between $ [-\pi,\pi]$. However, the imaginary $\mathcal{Q}''$ is different from zero for all frequencies. The dotted lines are a visual aid and indicate the position of the first forbidden band. 

To further analyze the dispersion relation, we divide our band structure in frequency intervals. 
The first interval defined by $I_{1}=\lbrace \omega: | \mathcal{Q}''d | = | c|  \rbrace$ where $c$ is a constant. In the context of electronic bands this constant must be zero meaning that $\mathcal{Q''} = 0$ which  defines an allowed band. However, for thermal bands $c \neq 0$. To understand this we point out that the real part $\mathcal{Q}'$ describes the propagating part of the thermal wave, while the imaginary part is related to its amplitude $|e^{i\mathcal{Q}x}|=|e^{-\mathcal{Q}''x}|$. The constant $c$ is therefore related to the damping of the wave as it travels the distance $d$. As $|\mathcal{Q}''d| = |c|<<1$ we get $|e^{i\mathcal{Q}x}|\approx |1-c|$ so the thermal wave damping is small and the signal manages to be transferred through the system. 

The following interval is $I_{2}=\lbrace \omega: $ for  $ \mathcal{Q}''>0$  $\frac{\partial \omega(\mathcal{Q})}{\partial \mathcal{Q}''}>0 \rbrace$ or the interval where the relation dispersion defines a convex curve. For energy bands this interval would define a band gap but in the thermal case it corresponds to a region where diffusive behavior is dominant as is shown in the reflectance with values close to unity. 

We want to emphasize, that unlike the quantum, photonic or acoustic cases, the band structure has allowed states in the frequency region corresponding to forbidden bands, determined by the cases when the Bloch vector takes imaginary values.

The role of the thermal bands is more clearly seen from the reflectance of the system. The calculated reflectance $R$  is shown in Figure \ref{fig:bandstructure}(b,d) and is consistent with the calculated thermal bands. The largest values of the reflectance correspond to largest values of $\mathcal{Q''}$ as we see in the interval $I_2$, indicating that attenuation due to $\mathcal{Q''}$  dominates over the wave-like behavior. From the wave perspective it means that the temperature signal is reflected by a high effective thermal impedance of the whole system that does not allow transmission. On the other hand, in regions with small values of $\mathcal{Q''}$ we note that  the wave-like behavior dominates over the diffusive, and the signal is allowed to travel through the crystal leading to small values of $R$. In the following intervals the behavior repeats itself.

Finally, we point out that in quantum mechanics, energy band structures can be seen as a result of the periodic interaction of atomic orbitals that make up the crystal. As a consequence, not all  electron wave functions in a crystal are allowed.  
 
However this explanation does not work for the thermal case and the appearance of bands is a consequence of the interference between thermal waves as well the attenuation due to $\mathcal{Q''}$ in the periodic system.  

\subsection{Finite Systems}
\begin{figure}[h]
\centering
\includegraphics[width=8cm,height=8cm]{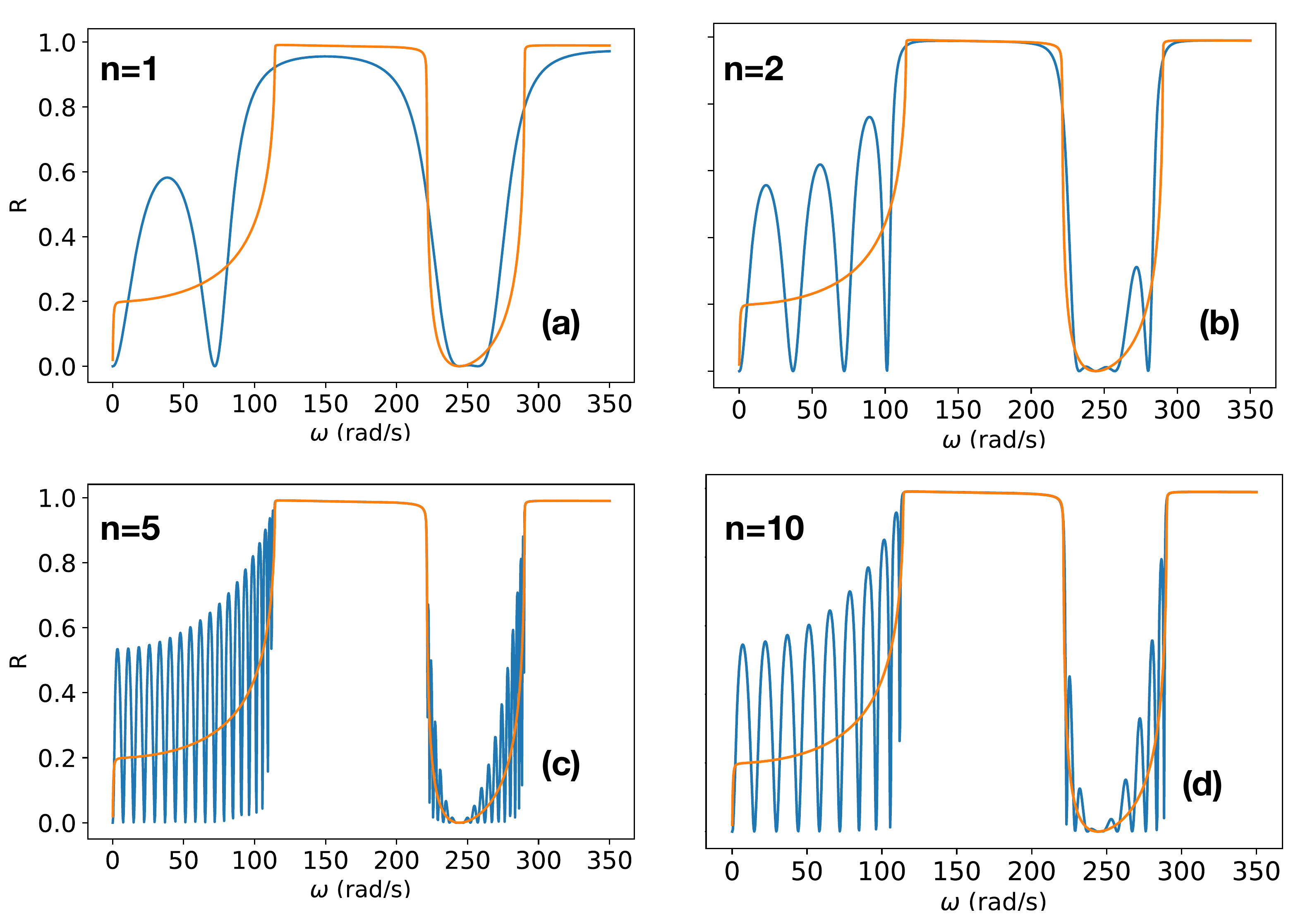}
\caption{Reflectance of finite layered system, for different number of bilayers $n$. (a) $n=1$, (b) $n=2$, (c) $n=5$, (d) $n=10$. The orange line is the reflectivity for a semi-infinite medium. For these calculations, $Z_0=Z_s=2Z_1$. }
\label{fig:layeredsystem2}
\end{figure}

Consider a system made of a finite number of bilayers, as shown in Figure \ref{fig:layeredsystem}(b). The bilayers are bounded by a substrate of impedance $Z_s$ and the incident medium $Z_0$. Each bilayer is made of layers $1-2$. In Figure \ref{fig:layeredsystem2} we show the reflectivity as a function of frequency for different number of bilayers (a) n=1, (b) n=3, (c) n=5, (d) n=10. As the number of layers increases, the regions of high reflectivity approaches the forbidden band of the semi-infinite system, indicated by the orange line in each panel. We see that even for $n=10$ bilayers the bandgaps are present. The oscillations outside the forbidden band are due to the multiple reflections from the boundaries of the structure.  The parameters of Fig.\ref{fig:layeredsystem2} are the same as those of Figure \ref{fig:bandstructure}(c,d). Thus, with a finite layered system, properties of the thermal crystal are obtained. 

\subsection{Thermal management}

\begin{figure}[h]
\centering
\includegraphics[width=8cm,height=8cm]{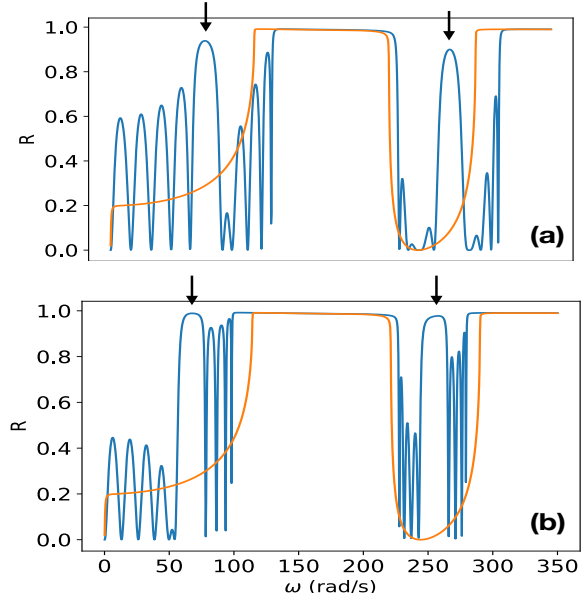}
\caption{The reflectivity of a system with $n=5$ bilayers with a defect included in every other layer bilayer.  (a) The defect is a layer with the same physical properties of layer $1$ but with a thickness of $.5d_1$. (b) The defect has the thickness $d_1$ but a thermal impedance $2Z_1$.  The introduction of defects adds additional forbidden states (indicated by the arrows).  As a reference, the orange line is the reflection for a semi-infinte crystal without defects.  }
\label{fig:layeredsystem2def}
\end{figure}

The temperature and heat flux can be further modified by the introduction of defects in an otherwise periodic system. Several cases are considered. For finite systems we introduce the defect by either changing the thickness of  a layer or its impedance.  As an example the defect is placed every-other bilayer. This is, a four bilayer system $1-2-1-2-1-2-1-2$ is changed to $1-2-D-2-1-2-D-2$. 
In this case the defect we have labeled layer $D$ and consider two possible defects. In one case the defective layer has the same physical properties as layer 1 but a different thickness $0.5d_1$. This is shown in  Figure \ref{fig:layeredsystem2def}(a)  where the reflectivity as a function of frequency is presented. In the second case,  Fig.\ref{fig:layeredsystem2def}(b) the thickness of the defect is $d_1$ but has an impedance $Z_D=2Z_1$. In both cases the defect introduces forbidden states indicated by the arrows outside the forbidden bands of the ordered system. 
The other parameters needed for the calculation are the same as those used in Figure \ref{fig:layeredsystem2}. 

\begin{figure}[h]
\centering
\includegraphics[width=8cm,height=9cm]{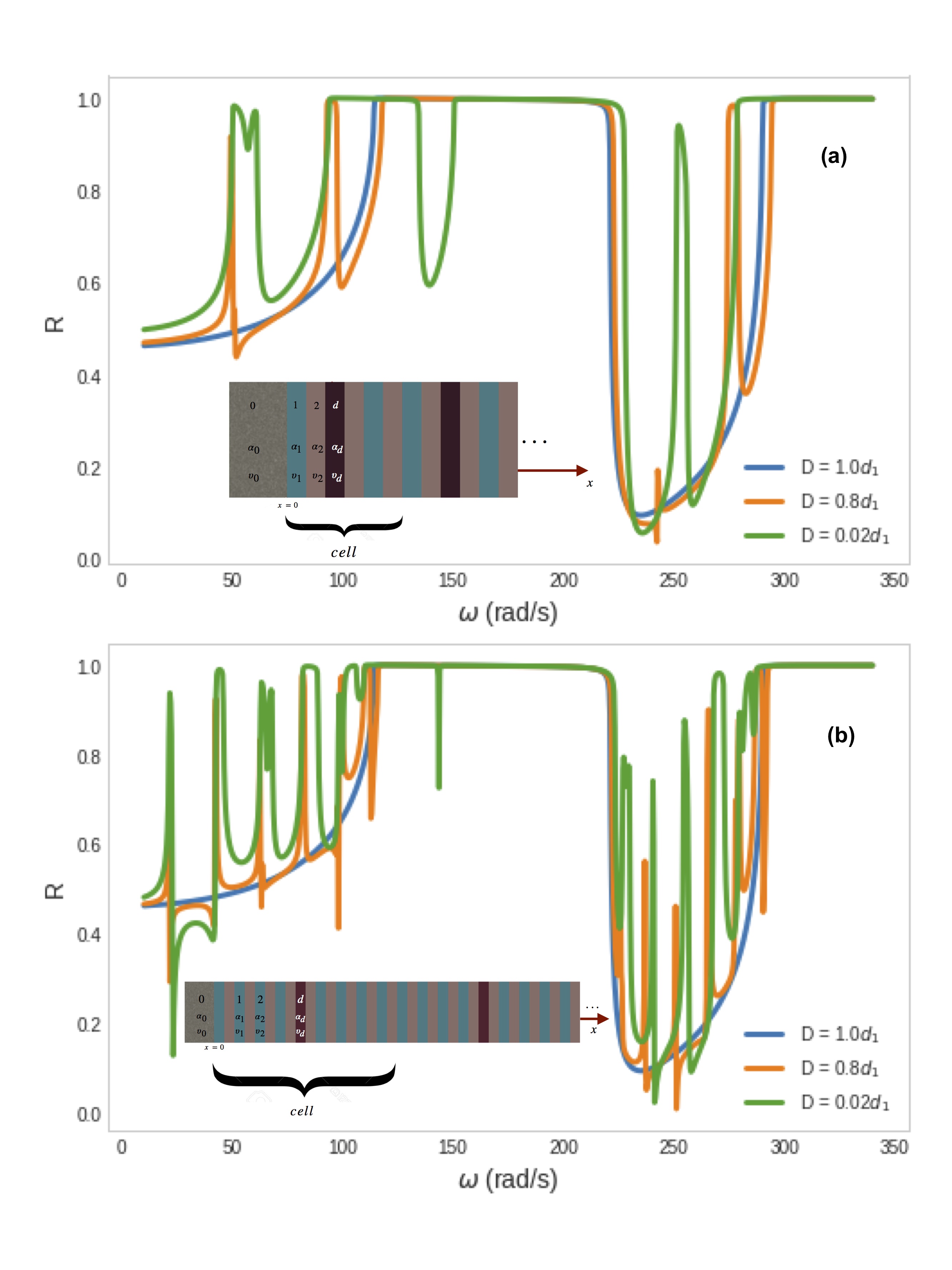}
\caption{The reflectivity of a thermal crystal with a defect in its unit cell. (a) The unit cell is made of 3 bilayers one with a defect. (b) The unit cell has 9 bilayers and a defect, as shown in the insets. In both caes, the different curves correspond to different defect thicknesses. For the blue line (no defect) $D=d_1$, for the orange line $D=0.8 d_1$ and for the green line $D=0.02 d_1$. }
\label{fig:crytaldef}
\end{figure}

Another way to add defect states is to build a thermal crystal (semi-infinite) where each unit cell has a defect. For example, let us assume a unit cell made of 3 bilayers. One of the individual layers has a different thickness (defect) $D$. The reflectivity is shown in Figure \ref{fig:crytaldef}(a). The inset shows where the defect is introduced. The three cases are for no defect $D=d_1$ (blue),  $D=0.8d_1$(orange) and  $D=0.02d_1$(green). Again, the appearance of allowed states within the forbidden band is observed as well as discrete forbidden states ($R\sim1$) in other frequency regions. 
  Similarly, in (b) we  show the case of a unit cell with 5 bilayers again introducing one defective layer. Most of the new states appear in the frequency regions corresponding to allowed bands. As the number of bilayers increases, the relative weight of the defect decreases. Let $N_D$ be the number of defective layers  and $N$ the total number of layers in each unit cell. We define the filling fraction of defects as $f=N_D/N$. If $f\rightarrow 0$, we recover the original system with no defects. In our case, for one defective layer $N_d=1$ and $f$ decreases as $N$ increases. For panel (a) we have $f=1/6$ and for (b) $f=1/18$. 
  The parameters of the layers are the same as Figure \ref{fig:bandstructure}(d).

\section{Conclusions}

In this work we presented a full analysis of the solution of the Cattaneo-Vernotte Equation for thermal waves in periodic systems.  Although the CVE is an hyperbolic equation with wave-like solutions for the temperature and heat flux, there are fundamental differences with other solutions of wave equations (acoustic, electromagnetic, Schr\"{o}dinger's equation). Firstly, the thermal waves do not carry energy \cite{Salazar_2006}, but this imposes no limitation on the wave behavior as far as reflection or  transmission properties.  The second difference is that even at normal incidence, the wave vector is complex for all frequencies as a result of the Floquet Theorem. Solution of the CVE in a periodic structure, leads to a dispersion relation that shows a pseudo-band structure. There are  frequency intervals where the reflectivity is close to one (forbidden bands) but do not satisfy the usual conditions of a forbidden band in periodic systems, thus the name pseudo-bands. 

In the case of finite periodic structures, we demonstrated that with no more that five layers we recover the main structure of a thermal crystal, recovering the forbidden regions in the reflectivity being able to make use of band engineering to control temperature and heat flow. Finally, following band engineering procedures, we show that the introduction of a defect can also be used to open allowed solutions within a forbidden region.  This means, in general,  that thermal band engineering is possible as well as designing thermal wave pass-filters. 

\section{Acknowledgements} D.B thanks the  support of Consiglio Nazionale delle Ricerche through BANDO ISM-BS002-2019 RM. This project was supported through Consejo Nacional de Ciencia y Tecnolog\'{i}a grant number A1-S-10537.

\section{Appendix A. Thermal waves}

Salazar \cite{Salazar_2006} has questioned some of the uses of thermal waves since he stresses that energy propagation is key to wave propagation. However, we differ from this point on view.
We adopt the following definition: \textit{a wave is any recognizable signal that is transferred from one part of a medium to another with a recognizable velocity of propagation} \citep{whitham}. In 1-D a signal is considered to be a wave if it is a solution to a PDE and is of the form $f(x-\upsilon t)$, where $f$ is a function of one variable and $\upsilon$ is a nonzero constant \citep{knobel}. 

To determine that solutions of the CVE are \textit{thermal waves} we need to determine the form of $f$ . Carrying out the change of variables $T(x,t)=U(\xi(x,t),\eta(x,t))$, assuming that the solution is given by $U(\xi,\eta)=U_{1}({\xi})U_{2}(\eta)$,  and substituting into the Eq.(\ref{eq:CV3}) we obtain

\begin{equation}
T(x,t)=e^{rt}f(x-\upsilon t)
\end{equation}

\noindent
where $r \in \mathbb{C}$ is known as the separation constant of the equation \citep{whitham,haberman}. Until now we know that thermal waves exist as mathematical objects, where the temperature signal travels with velocity $\upsilon$ modulated by a function that depends on time. The many tools of wave theory can therefore  be used  to study heat conduction in periodic systems.

\section{Appendix B. Floquet's Theorem}

Let the first order differential equation system $\frac{d}{dx}\vec{T}(x)=A\vec{T}(x)$, with the matrix  $A\in \mathbb{C}^{2\times 2}$ where has been imposed the periodicity $A(x+p)=A(x)$ for all $x\geq 0$, the fundamental matrix $F(x)=(\vec{T}_{1},\vec{T}_{2})$ is obtained using the Floquet Theorem, which establishes that

\begin{equation}
F(x)=Q(x)e^{Bx}
\label{eq:FloquetT1}
\end{equation}

\noindent
where $Q(x)\in C^{1}(\mathbb{R})$, has period $p$, is non-singular and $Q(0)=\mathbb{I}$; while the matrix $B\in \mathbb{C}^{2\times 2}$ satisfies the equation

\begin{equation}
F(p)=e^{Bp}.
\label{eq:FloquetT2}
\end{equation}

Given that $F(x)$ is a fundamental matrix, there must exists a non-singular matrix $C$ such that $F(x+p)=F(x)C$, which is named Monodromy Matrix \citep{folkersfloquet}. We evaluate this relation at $x=0$  and obtain that

\begin{equation}
C=F^{-1}(0)F(p)=e^{Bp}.
\end{equation}

We calculate the eigenvalues of monodromy matrix called Floquet's multipliers and denoted by $\lambda$; by another hand the eigenvalues of $B$, that we denote by $\mu$, are called Floquet's exponents. These eigenvalues are related between them by \citep{rolando1985}

\begin{equation}
\lambda=e^{\mu p}.
\end{equation}

This procedure is connected with the heat conduction problem developed before. We know that the solution given by Eq.\eqref{eq:matrixproblem}  is a fundamental matrix

\begin{equation}
F(x)=\left(\begin{array}{cc}
e^{ik_{j} x} &  e^{-i k_{j} x}\\
-iZ_{j}e^{ik_{j} x} & i Z_{j}e^{-i k_{j} x}\\
\end{array}\right),
\end{equation}

\noindent
therefore the monodromy matrix is

\begin{equation}
C=\left( \begin{array}{cc}
e^{ikp} & 0\\
0 & e^{-ikp}\\
\end{array}\right),
\end{equation}

\noindent
from $C$ we identify the Floquet exponents, $\mu=\pm ik$. When we repeat the unit cell $N$ times, the fundamental matrix is 

\begin{equation}
F(Np)=F((N-1)p+p)=\cdots=F(0)C^{N}.
\end{equation}


\bibliography{bandastermicas_v1}

\end{document}